\documentclass[preprint,aps,showkeys,showpacs,draft]{revtex4}
\usepackage[]{graphicx}
\begin{document}
\title{Anisotropic properties of mechanical characteristics and auxeticity of cubic crystalline media}
\author{T. Paszkiewicz}
\email{tapasz@prz.edu.pl} 
\author{S. Wolski}
\affiliation{Chair
of Physics, Rzesz\'{o}w University of Technology, Al. Powsta\'{n}c\'{o}w. Warszawy 6,
PL-35-959 Rzesz\'{o}w Poland} 

\begin{abstract}
Explicit expressions for inverse of Young's modulus $E(\textbf{n})$, inverse of shear modulus $G(\textbf{n},\textbf{m})$, and Poisson's ratio $\nu(\textbf{n},\textbf{m})$ for cubic media are considered. All these characteristics of elastic media depend on components $S_{11}$, $S_{12}$ and $S_{44}$ of the compliance tensor $\textbf{S}$, and on direction cosines of mutually perpendicular vectors $\textbf{n}$ and $\textbf{m}$ with fourfold symmetry axes. These characteristics are studied for \emph{all mechanically stable cubic materials} for vectors $\textbf{n}$ belonging to the irreducible body angle subtended by three cubic high symmetry directions $[001]$, $[111]$, and $[110]$. Regions in the stability triangle of in which cubic elastic materials are completely auxetic, nonauxetic, and auxetic are established. Several intermediate-valence compounds belonging to the region of complete auxecity are indicated. The extreme properties of $E^{-1}$, $G^{-1}$ and $\nu$ established by Hayes and Shuvalov are confirmed.
\end{abstract}

\maketitle

\section{Introduction}
Mechanical properties of elastic media are described by the bulk, Young's and the shear modules. Poisson's ratio is another important mechanical charactersitics of such materials. With the exception of bulk modulus, generally, these characteristics are anisotropic, i.e. they depend on direction of load and direction in which the response of a material is measured. The Poisson ratio is especially interesting, because it can be negative. Materials which exhibit a negative Poisson's ratio are referred to as auxetic media   \cite{baughman1,baughman2}. Auxetics respond to imposed uniaxial tension with lateral extension in place of expected contraction. Lakes described the synthesis of an actual auxetic material and proposed a simple mechanism underlying the negative Poisson's ratio \cite{lakes}. Such media can find interesting applications in future technologies. Ting and Barnett introduced classification of the auxetic behavior of anisotropic linear media \cite{ting}. 

Among the linear elastic media, a special attention has been devoted to cubic media \cite{baughman1,baughman2}. Turley and Sines used the technique of rotations by Euler angles \cite{turley}, and considered anisotropy of Young's modulus, shear modulus, and Poisson's ratio in cubic materials. Using the technique developed by Turley and Sines, Gunton and Saunders studied stability limits on the Poisson's ratio, and considered  martensitic transformations  \cite{gunton}. Milstein \cite{milstein1,milstein2}, Huang \emph{et al.} \cite{huang} analysed responses of ideal cubic crystals to uniaxial loadings.  

Poisson's ratio for cubic materials was studied by Milstein and Huang \cite{milstein3}, Jain and Verma \cite{jain}, Guo and Goddard \cite{guo}, Erdacos and Ren \cite{erdakos}, and Ting and Barnett \cite{ting}. With the exception of the last paper, the authors were looking for directions for which Poisson's ratio is negative. In the opinion of Baughman \emph{et al.}, negative Poissons ratios are a common feature of cubic media \cite{baughman1}. They refer specialy to cubic metal single crystals. Our results support this opinion \cite{paszkiewicz}.

An attempt to study more global properties of cubic auxetics was made by Tokmakova \cite{tokmakova}. To describe elastic properties of cubic media, she used the familiar elastic parameters $C_{11}/C_{44}$ and $C_{12}/C_{44}$. The purpose of our paper is to study anisotropy of all three mechanical characteristics of \emph{all mechanically stable elastic media}. To achieve this goal, we shall use different parameters, which were introduced by Every    \cite{every} and which were used by us for the unified description of elastic, acoustic and transport properties of cubic media    \cite{wilczynski1,wilczynski2,pruchnik1,pruchnik2}.  

\section{Elastic properties of cubic media}
\label{sc:elast-cubic}
Consider  homogeneous mechanically stable anisotropic elastic material. It is characterized by the components $S_{ijkl}\;(ij,k,l=1,2,3)$ of the compliance tensor $\textbf{S}$, or by the components $C_{ijkl}\;(ij,k,l=1,2,3) $ of the stiffness tensor $\textbf{C}$. Both these tensors have the structure $[[V^{2}]^{2}]$ \cite{s-s}, i.e. they have the complete Voigt's symmetry, e.g. $C_{ijkl}=C_{ji,kl}=C_{kl,ij}$. The compliance and stiffness tensors are mutually inverse, i.e. 
\begin{equation}
\textbf{S}=\textbf{C}^{-1}. 
\label{eq:S-inv-of-C}
\end{equation}

After Walpole \cite{walpole}, we denote degenerate eigenvalues of \textbf{C} by $c_{J}$, $c_{L}$, and $c_{M}$. One has \cite{pawol}
\begin{eqnarray}
c_{J}=C_{11}+2C_{12},\:	c_{L}=2C_{44},\: c_{M}=C_{11}-C_{12},\nonumber  \\ 
s_{J}=c_{J}^{-1}, \: s_{L}=c_{L}^{-1}, \: s_{M}=c_{M}^{-1}. 
\label{eq:c-and-s}
\end{eqnarray}
Similarly $s_{J}$, $s_{L}$, and $s_{M}$ are eigenvalues of $\textbf{S}$. 

The eigenvalues $s_{J}$, $s_{L}$, and $s_{M}$ can be written in terms of $S_{11}$, $S_{12}$ and $S_{44}$, namely, 
\begin{equation}
	s_{J}=S_{11}+2S_{12}, \, s_{L}=S_{44}/2,\, s_{M}=S_{11}-S_{12}.
\label{eq:w2v}
\end{equation}
Using these relations and the second line of Eqs. (\ref{eq:c-and-s}), we express the matrix elements $S_{UV}$ by $C_{UV}$ ($U,V=1,2,4$)
\begin{equation}
	S_{11}=\frac{C_{11}+C_{12}}{c_{J}c_{M}},\, 	  		S_{12}=-\frac{C_{12}}{c_{J}c_{M}},\, S_{44}=\frac{1}{C_{44}}.
\end{equation}
These expressions are in agreement with the familiar results \cite{nye}. 

One may use eigenvalues $c_{J}$, $c_{L}$, and $c_{M}$ in place of three elastic constants $C_{11}$, $C_{12}$, and $C_{44}$, but the more reasonable choice is the use of three parameters introduced by Every \cite{every}, namely, $s_{1}=\left(C_{11}+2C_{44}\right)$ and two dimensionless parameters
\begin{equation}
	s_{2}=\frac{\left(C_{11}-C_{44}\right)}{s_{1}}, s_{3}=\frac{\left(C_{11}-C_{12}-2C_{44}\right)}{s_{1}}.
\end{equation}
The inverse relations hold \cite{pruchnik2}
\begin{equation}
	C_{11}=s_{1}\left(2s_{2}+1\right)/3,\; C_{12}=s_{1}\left(4s_{2}-3s_{3}-1\right)/3,\; C_{44}=s_{1}\left(1-s_{2}\right)/3.
\label{eq:inverse}
\end{equation}
The elements of \textbf{C} and eigenvalues $c_{U}\, (U=J,L,M)$ are proportional to $s_{1}$. From Eq. (\ref{eq:S-inv-of-C}), it follows that $S_{ij}\propto s^{-1}_{1}$. Therefore, we introduce the dimensionless components of the compliance tensor $S_{ij}'=s_{1}S_{ij}$.

\begin{figure}[htpb]
	\centering
		\includegraphics[bb=0 590 165 842, clip]{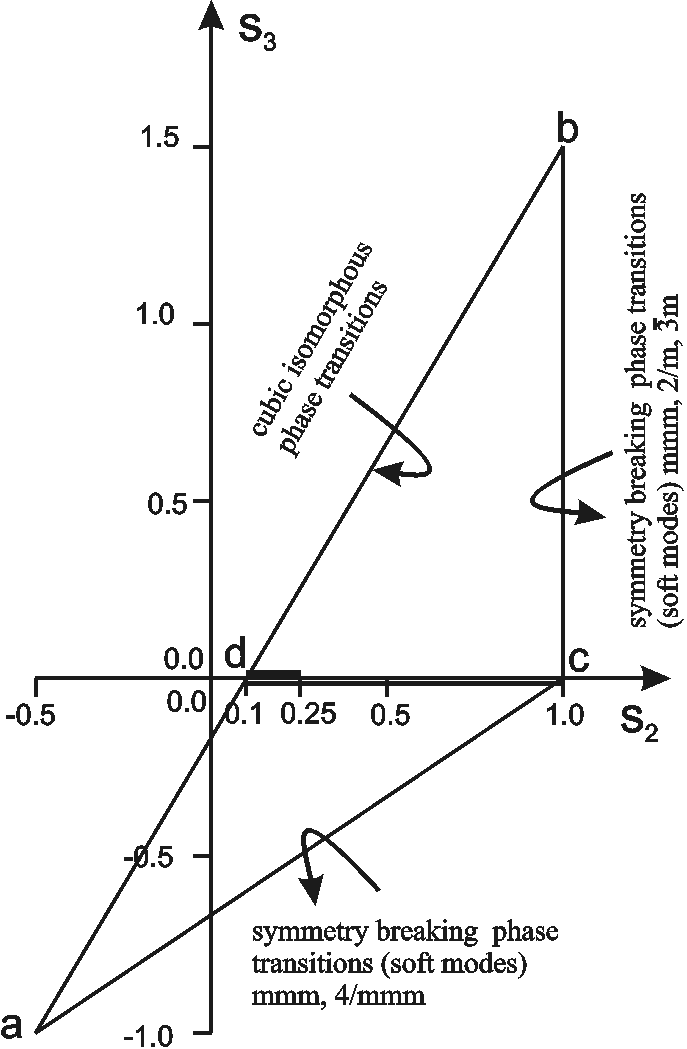}
	\caption{Triangle abc of stability (ST) of cubic media. Points $\left(s_{1}, s_{2}, s_{3}\right)$ characterizing mechanically stable cubic materials belong to the prism ($s_{3}>0$), the base of which is the stability triangle.}
	\label{fig:Trojstab}
\end{figure}

The mechanical stability is guaranteed when all introduced eigenvalues are positive. This condition yields familiar inequalities \cite{wallace}
\begin{equation}
\label{eq:inequalities}
\left(A_{11}+2A_{12}\right)>0, \: \left(A_{11}-A_{12}\right)>0, \: A_{44}>0,\; (A=C,\, S).	
\end{equation}
In terms of Every's parameters, cubic material is stable, if $s_{1}>0$ and $s_{2}$, $s_{3}$ belong to the \emph{stability triangle} (ST for short) in the plane ($s_{2}$, $s_{3}$) (cf. Fig. \ref{fig:Trojstab})
\begin{equation}
	\left(1+2s_{2}\right)>\left|4s_{2}-3s_{3}-1\right|,\; \left(10s_{2}-6s_{3}\right)>1, \;	s_{2}<1. 
	\nonumber
\end{equation}

Introduce two parameters of elastic anisotropy
\begin{eqnarray}
\chi_{c}\equiv \left(c_{M}-c_{L}\right)=\left(C_{11}-C_{12}-2C_{44}\right)=s_{1}s_{3},\nonumber  \\ 
\chi_{s}\equiv s_{1}\left(s_{M}-s_{L}\right)=-s_{1}\left(\frac{1}{c_{M}}-\frac{1}{c_{L}}\right)=-\frac{s_{1}\chi_{c}}{c_{L}c_{M}}=-\frac{s_{1}^{2}s_{3}}{c_{L}c_{M}}.	
\label{eq:chi's}
\end{eqnarray}
The parameter $\chi_{s}$ is dimensionless.  

For $s_{3}=0$ both $\chi_{c}$, and $\chi_{s}$ vanish, hence the  anisotropic parts od $E^{-1}$, $\nu$ and $G^{-1}$ vanish (cf. Sect. \ref{sc:anisotropy}). In this case the obtained expressions for mechanical characteristics are the same as for isotropic materials (cf.\cite{pawol}). Therefore the interval $dc$ ($1/10<s_{2}<1$, $s_3=0$), indicated by the line lying below the $s_2$-axis, is the interval of isotropic materials. We shall demonstrate later that the \emph{isotropic auxetic materials of cubic symmetry} are located on the interval $1/10<s_{2}<1/4\; (s_{3}=0)$ indicated by the interval lying above the $s_2$-axis. 
\section{Mechanical characteristics of elastic media}
\label{sc:mechanical}
\begin{figure}[htpb]
	\centering
		\includegraphics[bb=0 710 171 842, clip]{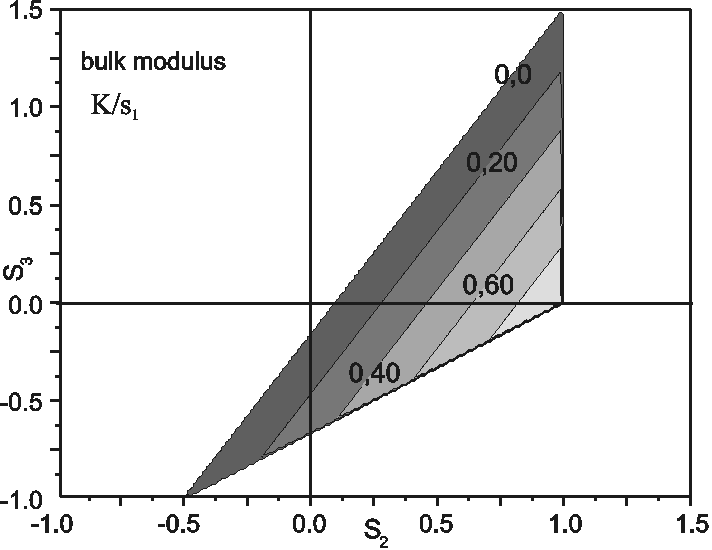}
	\caption{Contour map of dimensionless bulk modulus $K\left(s_{2},s_{3}\right)/s_{1}$}
	\label{fig:bulk-modulus}
\end{figure}

The response of elastic materials to hydrostatic pressure is characterized by the bulk modulus $K$ \cite{nye}. For cubic elastic media
\begin{equation}
	K=\left[3\left(S_{11}+2S_{12}\right)\right]^{-1}=\left(3s_{J}\right)^{-1}=\left(C_{11}+2C_{12}\right)/3. 
\label{eq:bulk}
\end{equation}
In Fig. \ref{fig:bulk-modulus} we present the dependence of dimensionless bulk modulus $K/s_{1}$ on the parameters $s_{2}$, $s_{3}$

We shall characterize the response of the elastic material to uniaxial tension in a direction $\textbf{n}\; (\textbf{n}\textbf{n}=1)$ by the dimensionless Young's modulus $e(\textbf{n})=E(\textbf{n})/s_{1}$
\begin{equation}
	e^{-1}(\textbf{n})=(\textbf{n}\otimes\textbf{n})\cdot\textbf{S}'\cdot(\textbf{n}\otimes\textbf{n})=n_{i}n_{j}S_{ijkl}'n_{k}n_{l}.
\end{equation}
Repetition of a suffix in a product of tensors or in a single tensor implies the usual summation with the respect to that suffix over the values 1,2,3.

Similarly, the response to the shear stress will be characterized by the dimensionless shear modulus $g(\textbf{n},\textbf{m})=G(\textbf{n},\textbf{m})/s_{1}$, ($\textbf{m}\textbf{m}=1,\;\textbf{n}\textbf{m}=0$)
\begin{equation}
\frac{1}{4g(\textbf{m},\textbf{n})}=(\textbf{m}\otimes\textbf{n})\cdot S'\cdot(\textbf{m}\otimes\textbf{n})=m_{i}n_{j}S_{ijkl}'m_{k}n_{l}.  
\label{eq:shear}
\end{equation}
Poisson's ratio $\nu(\textbf{n},\textbf{m})$ is the ratio of the lateral contraction along direction $\textbf{m}$ to the longitudinal extension along  direction $\textbf{n}$ perpendicular to $\textbf{m}$
\begin{equation}
\nu(\textbf{m},\textbf{n})=-e(\textbf{n})(\textbf{m}\otimes\textbf{m})\cdot S'\cdot(\textbf{n}\otimes\textbf{n}) = -\frac{m_{i}m_{j}S_{ijkl}'n_{k}n_{l}}{n_{t}n_{u}S_{tuvw}'n_{v}n_{w}}. 
\label{eq:poisson}
\end{equation}

Using the basis of tensors $[[V^{2}]^{2}]$ constructed for cubic media by Walpole \cite{walpole}, we obtained the explicit expressions for $E(\textbf{n})$, $\left[4G(\textbf{m},\textbf{n})\right]^{-1}$, and $\left[-\nu(\textbf{m},\textbf{n})/E(\textbf{n})\right]$ for media of high and middle symmetry classes \cite{pawol}. For cubic materials, we obtained the familiar Nye \cite{nye} and Ting and Barnett formulae \cite{ting}
\begin{eqnarray}
E^{-1}(\textbf{n})=\left[\left(s_{J}-s_{M}\right)/3+s_{L}\right]+\left(s_{M}-s_{L}\right)p(\textbf{n})= \nonumber \\
S_{11}-2\left(S_{11}-S_{12}-S_{44}/2\right)T(\textbf{n})=S_{11}\left[1-3T(\textbf{n})\right]+\left(S_{11}+2S_{12}+S_{44}\right)T(\textbf{n}),
\label{young-cub}
\end{eqnarray}
where $p\left(\textbf{n}\right)=\sum_{i=a}^{c}n_{i}^{4}$, and $T(\textbf{n})=\left(n_{a}^{2}n_{b}^{2}+n_{b}^{2}n_{b}^{2}+n_{c}^{2}n_{a}^{2}\right)$. We also denoted the fourfold axes of cubic media by $\textbf{a}$, $\textbf{b}$, and $\textbf{c}$ and, for example, $m_{a}=(\textbf{m}\textbf{a})$.

Similarly, for $\nu(\textbf{m},\textbf{n})$ and $G(\textbf{m},\textbf{n})$ we get
\begin{eqnarray}	
-\frac{\nu(\textbf{m},\textbf{n})}{E(\textbf{n})}=\left(s_{J}-s_{M}\right)/3+\left(s_{M}-s_{L}\right)P(\textbf{m},\textbf{n}),\\ \label{nu-cub}		
\left[4G(\textbf{m},\textbf{n})\right]^{-1}=s_{L}/2+\left(s_{M}-s_{L}\right)P(\textbf{m},\textbf{n}). 
	\label{g-cub}
\end{eqnarray}
The function $P\left(\textbf{m},\textbf{n}\right)$ was introduced by Every \cite{every} (cf. also Ting and Barnett \cite{ting}) 
\begin{equation}
P\left(\textbf{m},\textbf{n}\right)=P\left(\textbf{n},\textbf{m}\right) =\sum_{i=a}^{c}\left(m_{i}n_{i}\right)^{2}.
\label{eq:P-function}
\end{equation}
Introduced dimensionless mechanical characteristics of elastic materials are functions of \emph{five} variables. This fact complicates studies of their properties. 

The functions $p(\textbf{n})$ and $P(\textbf{m},\textbf{n})$ obey the familar inequalities 
\begin{eqnarray}
	1/3=p\left(\left\langle111\right\rangle\right)\leq p\left(\textbf{n}\right)\leq 1=p\left(\left\langle001\right\rangle\right), \label{eq:p-ineq} \\ 0=P\left(\left\langle100\right\rangle,\:\left\langle001\right\rangle\right)\leq P\left(\textbf{m},\textbf{n}\right)\leq  1/2=P\left(\left\langle-110\right\rangle,\:\left\langle110\right\rangle\right). \label{eq:P-ineq}
	\end{eqnarray}

Note that the first term of expression defining $E^{-1}(\textbf{n})$ and  $\left[-\nu(\textbf{m},\textbf{n})/E({\textbf{n}})\right]$ is isotropic, whereas the second one is anisotropic. Shear modulus has the same structure. 

In the Appendix we write down the explicit formulas for three considered anisotropic mechanical characteristics in terms of components of both stiffness and compliance tensors and consider the limit of isotropic elastic materials. 
\section{Anisotropy of mechanical characteristics of elastic materials}
\label{sc:anisotropy}
Using Eq. (\ref{eq:w2v}) expressions for $1/e(\textbf{n})$, $1/g(\textbf{m},\textbf{n})$ and $\nu(\textbf{m},\textbf{n})$ can be cast to the form used by Hayes and Shuvalov \cite{hayes} 
\begin{eqnarray}
e^{-1}(\textbf{n})=S_{12}'+S_{44}'/2+\chi_{s}\sum_{i=1}^{3}n_{i}^{4},\\ \label{eq:inv-E}
	g^{-1}(\textbf{m},\textbf{n})=S_{44}'+4\chi_{s}\sum_{i=1}^{3}m_{i}^{2}n_{i}^{2},\\ \label{inv-G}	\nu(\textbf{m},\textbf{n})=-\frac{\Gamma\left(\textbf{m},\textbf{n};s_{2},s_{3}\right)}{S_{12}'+S_{44}'/2+\chi_{s}\sum_{j=1}^{3}n_{j}^{4}},
\label{nu}
\end{eqnarray}
where $\Gamma\left(\textbf{m},\textbf{n};s_{2},s_{3}\right)=S_{12}'\left(s_{2},s_{3}\right)+\chi_{s}\left(s_{2},s_{3}\right)\sum_{i=1}^{3}m_{j}^{2}n_{i}^{2}$.

The dimensionless quantities -- Young's modulus, shear modulus and Poisson's ratio -- are functions of angles and Every's parameters $s_{2}$, $s_{3}$. Studying their angular dependence, we may confine ourselves to the body angle $\Omega_{c}$ subtended by three vectors of high symmetry $\textbf{n}_{1}=[001]$, $\textbf{n}_{2}=[111]$, $\textbf{n}_{3}=[110]$.  An arbitrary unit vector $\bar{\textbf{n}}$ belonging to $\Omega_{c}$ is a linear combination of these three vectors 
\begin{equation}
	\bar{\textbf{n}}=\alpha \textbf{n}_{1}+\beta\textbf{n}_{2}+\gamma\textbf{n}_{3}\; (\bar{\textbf{n}}\bar{\textbf{n}}=1).	
\end{equation}
All unit vectors belonging to the body angle $4\pi$ can be obtained as a result of the application of cubic symmetry operations to vectors belonging to $\Omega_{c}$. 

\begin{figure}[htpb]
	\centering
		\includegraphics[bb=1 594 316 842, clip]{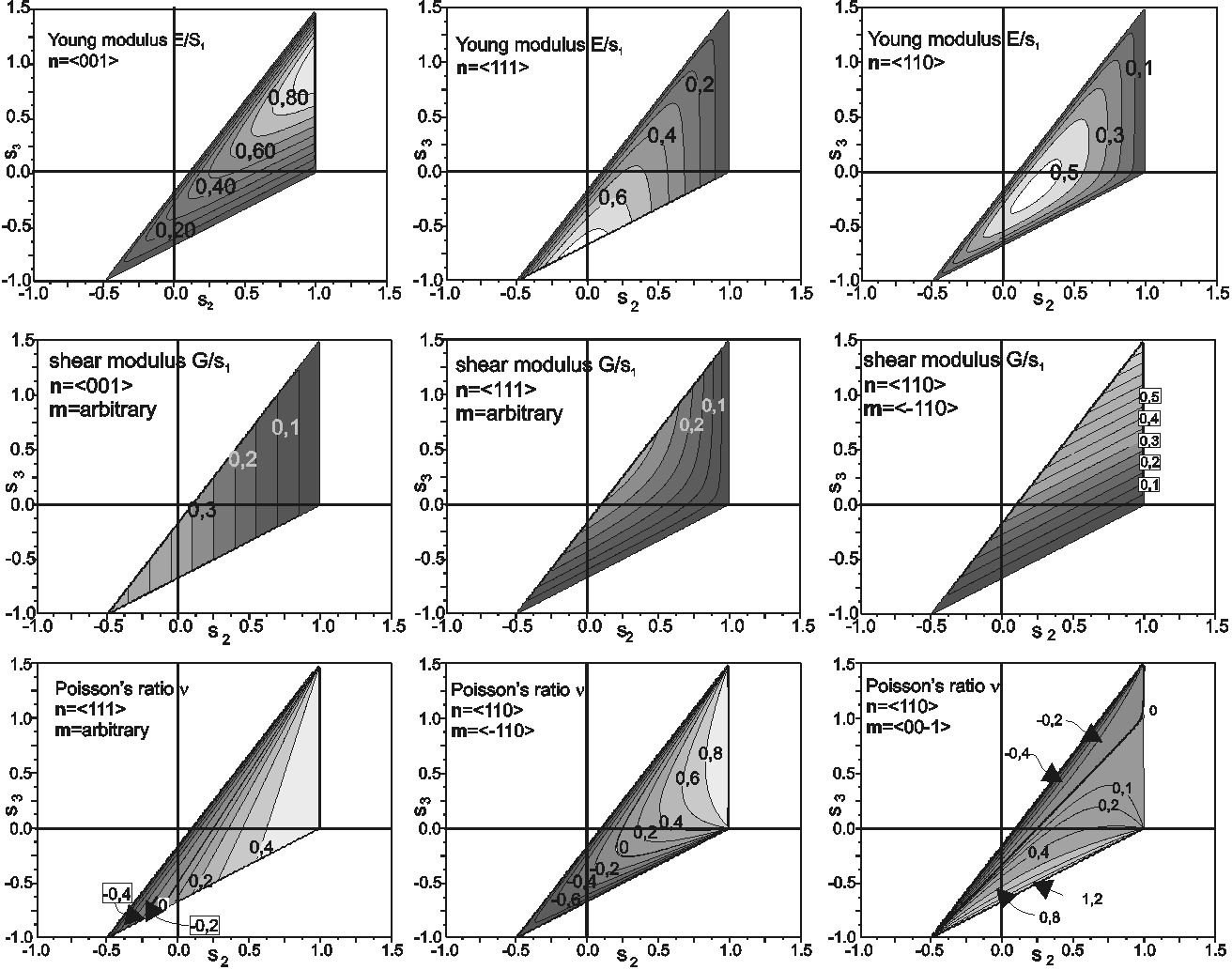}
	\caption{Anisotropy of mechanical characteristics of elastic media: contour maps of $e\left(\textbf{n},s_{2},s_{3}\right)$, $g\left(\textbf{n},s_{2},s_{3}\right)$, and $\nu\left(\textbf{n},s_{2},s_{3}\right)$ for $\textbf{n}=\textbf{n}_{1}=\left\langle 001\right\rangle$, $\textbf{n}=\textbf{n}_{2}=\left\langle 111\right\rangle$ and $\textbf{n}=\textbf{n}_{3}=\left\langle 110\right\rangle$}
	\label{fig:e-g-nu-n1-n3}
\end{figure}

Contour maps of $e\left(\textbf{n},s_{2},s_{3}\right)$, $g\left(\textbf{n},s_{2},s_{3}\right)$, and $\nu\left(\textbf{m},\textbf{n},s_{2},s_{3}\right)$ for vectors $\textbf{n}_{1}$, $\textbf{n}_{2}$, and $\textbf{n}_{3}$ are shown in Fig. \ref{fig:e-g-nu-n1-n3}. 

\section{Inequalities for elastic characteristics of cubic elastic media}
\label{sc:inequalities}
Properties of mechanical characteristics of elastic media depend on the sign of $\chi_{s}$. In the lower part of ST (where $s_{3}<0$), $\chi_{s}>0$, whereas on the upper part of ST, ($s_{3}>0$) $\chi_{s}<0$. 

\begin{figure}[htpb]
	\centering
		\includegraphics[bb=0 710 171 842, clip]{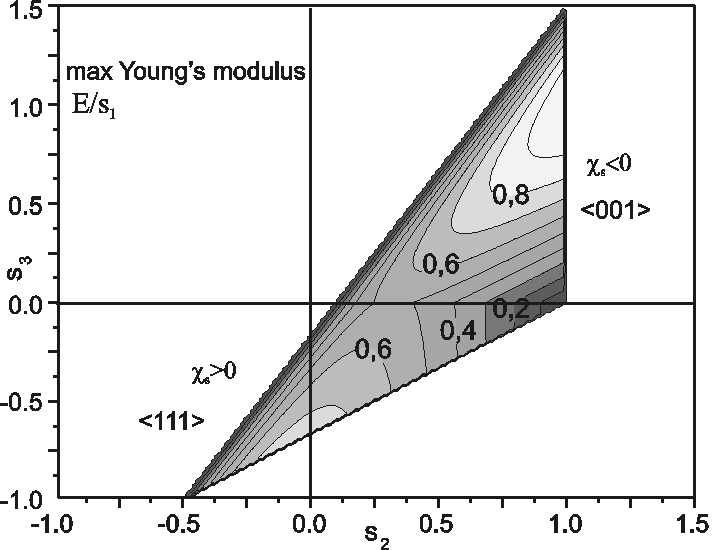}
	\caption{Contour map of the surface of local maxima of dimensionless Young's modulus defined by the function $F^{(e)}_{max}\equiv F^{(e)}_{1}\left(s_{2},s_{3} \right)\theta(\chi_{s})+F^{(e)}_{2}\left(s_{2},s_{3} \right)\theta(-\chi_{s})$.}
	\label{fig:Fe-on-isotropy-line}
\end{figure}

Hayes and Shuvalov studied extreme properties of mechanical characteristics of cubic elastic media. Let us summarize their results \cite{hayes}. If $\chi_{s}>0$, then at each point of ST the function $e\left(\textbf{n},s_{2},s_{3}\right)$ reaches a local maximum $e^{(+)}_{max}$ for $\textbf{n}=\left\langle 111\right\rangle$, while for $\textbf{n}=<001>$ at each point of ST it attains a local minimum $e^{(+)}_{min}$
\begin{eqnarray}
	e_{max}^{(+)}\left(s_{2},s_{3}\right)=\frac{3}{S_{11}'+2S_{12}'+S_{44}'}\equiv F^{(e)}_{1}\left(s_{2},s_{3}\right)\; \left(\textbf{n}=\left\langle 111\right\rangle \right),\nonumber\\ 
  e_{min}^{(+)}\left(s_{2},s_{3}\right)= \frac{1}{S_{11}'}\equiv F^{(e)}_{2}\left(s_{2},s_{3}\right)\; \left(\textbf{n}=\left\langle 001\right\rangle \right).
	\nonumber
\end{eqnarray}
If $\chi_{s}>0$, at each point of ST, value of $e$ for $\bar{\textbf{n}}\in\Omega_{c}$ obeys the inequality $e_{min}^{(+)}\leq e^{(+)}(\bar{\textbf{n}})\leq e_{max}^{(+)}$. Reliefs defined by values of functions $F^{(e)}_{1}\left(s_{2},s_{3}\right)$ and $F^{(e)}_{2}\left(s_{2},s_{3}\right)$ intersect along the isotropy interval ($s_{3}=0$; $1/10<s_{2}<1$), where neither $e$ nor $g$ (nor $\nu$) do not depend on angles (cf. Fig. \ref{fig:Fe-on-isotropy-line}). 

If $\chi_{s}<0$, then
\begin{eqnarray}
e_{max}^{(-)}\left(s_{2},s_{3}\right)=F^{(e)}_{2}(s_{2},s_{3}),\, \left(\textbf{n}=\left\langle 001\right\rangle \right),\nonumber \\ 
 e_{min}^{(-)}\left(s_{2},s_{3}\right)=F^{(e)}_{1}(s_{2},s_{3}), \left(\textbf{n}=\left\langle 111\right\rangle \right). \nonumber 		
\end{eqnarray}
Relief for local minima of $e\left(\textbf{n},s_{2},s_{3} \right)$ is obtained with the help of function $F^{(e)}_{min}\equiv F^{(e)}_{2}\left(s_{2},s_{3}\right)\theta(\chi_{s})+F^{(e)}_{1}\left(s_{2},s_{3}\right)\theta(-\chi_{s})$, where $\theta(x)$ is the Heaviside step function.

\begin{figure}[htpb]
	\centering
		\includegraphics[bb=0 709 171 842, clip]{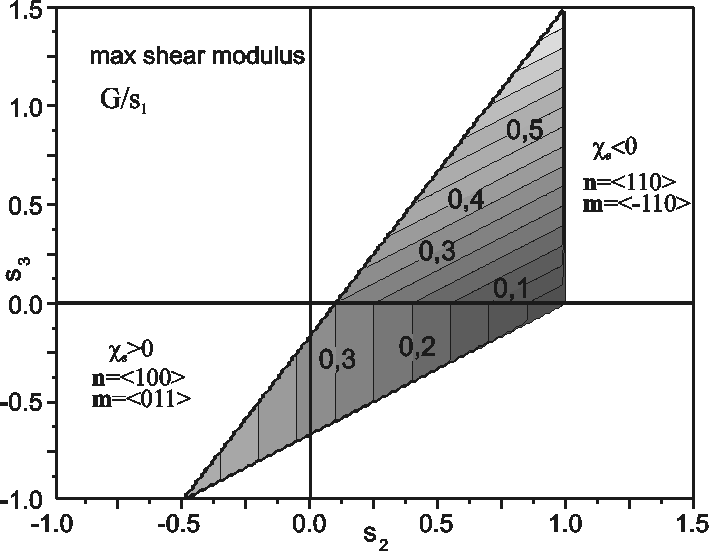}
	\caption{Contour map of the surface of local maxima of dimensionless shear modulus defined by function $F^{(g)}_{1}\left(s_{2},s_{3} \right)\theta(\chi_{s})+F^{(g)}_{2}\left(s_{2},s_{3} \right)\theta(-\chi_{s})$. }
	\label{fig:Fg-on-isotropy-line}
\end{figure}

Consider extreme properties of dimensionless shear modulus. 
If $\chi_{s}>0$, then 
\begin{eqnarray}		
	g_{max}^{(+)}=\frac{1}{S_{44}'}\equiv F^{(g)}_{1}\left(s_{2},s_{3}\right)\,\left(\textbf{m}=\left\langle 100 \right\rangle, \textbf{n}=\left\langle 011\right\rangle; \textbf{n}=\left\langle 100 \right\rangle, \textbf{m}=\langle 011\rangle \right)\nonumber \\ 
	g_{min}^{(+)}=\frac{1}{2\left(S_{11}'-S_{12}' \right)}\equiv F^{(g)}_{2}\left(s_{2},s_{3}\right) \,\left(\textbf{m}=\left\langle 110 \right\rangle, \textbf{n}=\left\langle -110\right\rangle; \textbf{n}=\left\langle 110 \right\rangle, \textbf{m}=\langle -110\rangle \right).\nonumber 
\end{eqnarray}
If $\chi_{s}<0$, then 
\begin{eqnarray}	
	g_{max}^{(-)}=F^{(g)}_{2}\left(s_{2},s_{3}\right) , \,\left(\textbf{m}=\left\langle -110 \right\rangle, \textbf{n}=\left\langle 110\right\rangle; \textbf{n}=\left\langle -110 \right\rangle, \textbf{m}=\langle 110\rangle \right)\nonumber \\  
	g_{min}^{(-)}=F^{(g)}_{1}\left(s_{2},s_{3}\right), \,\left(\textbf{m}=\left\langle 100 \right\rangle, \textbf{n}=\left\langle 011\right\rangle; \textbf{n}=\left\langle 100 \right\rangle, \textbf{m}=\langle 011\rangle \right)\nonumber. 
\end{eqnarray}	
Properties of surface of local maxima of the dimensionless shear modulus are summarized in Fig. \ref{fig:Fg-on-isotropy-line}

\begin{figure}[htpb]
	\centering
		\includegraphics[bb=0 709 171 842, clip]{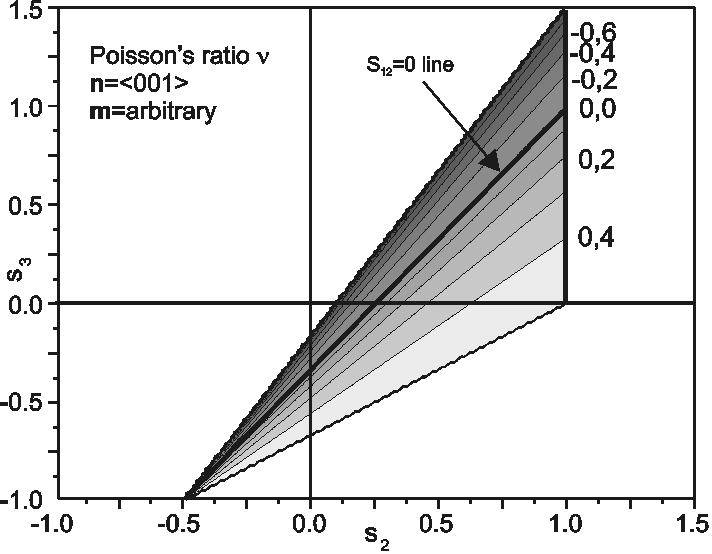}
	\caption{Contour map of relief of $F_{\nu}\left(s_{2},s_{3}\right)$. In the upper part of ST, it is related to geometrical loci of local maxima of $\nu$. In the lower part of ST, the contour map is related to geometrical loci of local minima of $\nu$. }
	\label{fig:F-nu}
\end{figure}

In the case of $\nu\left(\textbf{m},\textbf{n};s_{2},s_{3}\right)$, Hayes and Shuvalov \cite{hayes} have proved that if $\chi_{s}>0$, at each point of the lower part of ST, the Poisson ratio attains a local maximum for $\textbf{m}=\left\langle011 \right\rangle, \textbf{n}=\left\langle100 \right\rangle$, whereas for $\textbf{m}=\left\langle 011\right\rangle, \textbf{n}=\left\langle 100\right\rangle$ at each point of the upper part of ST it attains a local minimum
\begin{eqnarray}
	\nu_{max}\left(s_{2},s_{3}\right)=-\frac{S_{12}'}{S_{11}'}\equiv F_{\nu}\left(s_{2},s_{3}\right)\, \left(\textbf{m}=\left\langle011 \right\rangle, \textbf{n}=\left\langle100 \right\rangle,\: \chi_{s}>0\right),\nonumber \\
	\nu_{min}\left(s_{2},s_{3}\right)=F_{\nu}\left(s_{2},s_{3}\right), \,\left(\textbf{m}=\left\langle 011\right\rangle, \textbf{n}=\left\langle 100\right\rangle  \right),\: \chi_{s}<0 ).
	\label{fig:HS-ineq}
\end{eqnarray}
Contour map of relief of the function $F_{\nu}\left(s_{2},s_{3}\right)$ is depicted in Fig. \ref{fig:F-nu}.

In virtue of inequalities (\ref{eq:inequalities}), the extreme values of $\nu(\textbf{m},\textbf{n})$ satisfies the inequalities \cite{hayes} (cf. Fig. \ref{fig:F-nu})
\begin{eqnarray}
{-1} \leq \nu_{min} \leq 0 \;  \left(\chi_{s} >0,\;  S_{12}>0 \right), \nonumber \\
{0} \leq \nu_{max} \leq 1/2\;  \left( \chi_{s} <0,\; S_{12}<0 \right).\nonumber
\end{eqnarray}
\section{Properties of Poisson's ratio}
\label{sc:poisson}
In this section we use the results obtained by Ting and Barnett \cite{ting} and relate them to regions of the stability triangle. The relation (\ref{nu-cub}) can be written in the form used by Ting and Barnett \cite{ting}
\begin{equation}
-\frac{\nu(\textbf{m},\textbf{n};s_{2},s_{3})}{e(\textbf{n};s_{2},s_{3})}=\left[1-2P(\textbf{m},\textbf{n})\right]S_{12}'(s_{2},s_{3})+Q(s_{2},s_{3})P(\textbf{m},\textbf{n})/2.	
\nonumber
\end{equation}
with $Q=\left[2\left(S_{11}'+S_{12}'\right)-S_{44}'\right]$.

\begin{figure}[htpb]
	\centering
		\includegraphics[bb=0 620 397 842, clip]{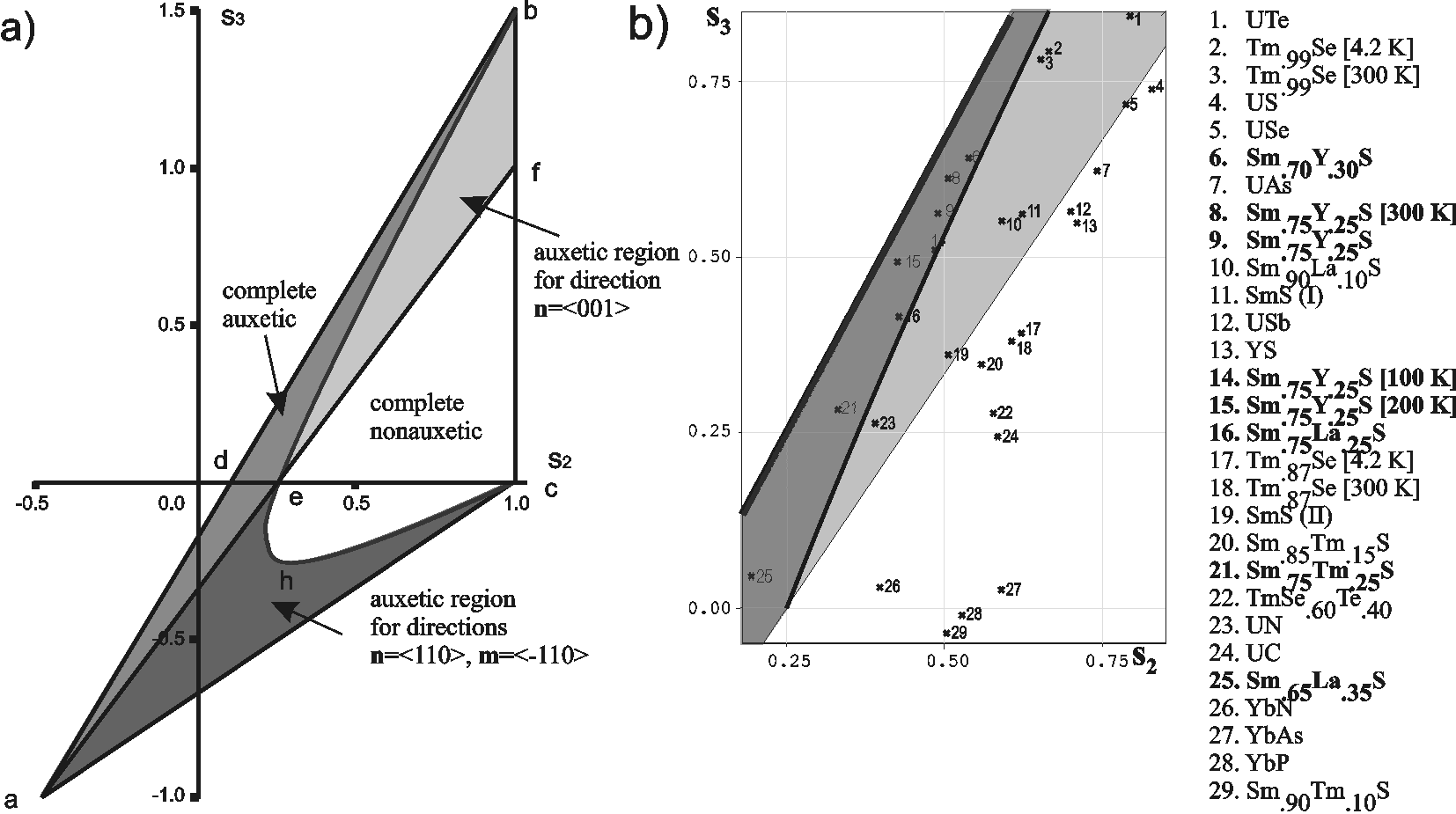}
	\caption{Auxetic properties of cubic elastic media -- characteristic regions of the stability triangle. Insert: the enlarged part of ST with some intermediate-valence compounds indicated. Compounds distinguished by bold letters are the completely auxetics (cf. \cite{pruchnik2}).}
	\label{fig:regions-ST}
\end{figure}

I. Consider a pair of unit vectors: $\textbf{m}=\left\langle-110\right\rangle$ and $\textbf{n}=\left\langle110\right\rangle$. For it,  $P(\textbf{m},\textbf{n})$ attains the maximal value $1/2$, and $\nu=Q/4$.\newline  
i) If $S_{12}'\left(s_{2},s_{3}\right)>0$, $Q\left(s_{2},s_{3}\right)>0$, then $\nu/e\leq-Q/4$. For $s_{3}>0$, the inequalities imposed on $S_{12}'$ and $Q$ define  the region $deb$, whereas for $s_{3}<0$ -- the region $aed$ of ST (cf. Fig. \ref{fig:regions-ST}). Because for all pairs $(\textbf{m},\textbf{n})$ Poisson's ratio is negative, elastic materials belonging to these two regions are \emph{completely} auxetics.\newline
ii) If $S_{12}'\left(s_{2},s_{3}\right)<0$, $Q\left(s_{2},s_{3}\right)<0$, then $\nu/e\geq-Q/4$. For $s_{3}>0$, the inequalities imposed on $S_{12}'$ and $Q$ define the region $cfe$ , whereas for $s_{3}<0$ -- the region $ceh$ of ST (Fig. \ref{fig:regions-ST}). Because for all pairs $(\textbf{m},\textbf{n})$ Poisson's ratio is positive, elastic materials belonging to these two regions are \emph{non} auxetics.\newline   
iii) If $S_{12}'\left(s_{2},s_{3}\right)>0$, $Q\left(s_{2},s_{3}\right)<0$, then $\nu/e\leq -Q/4$. The inequalities imposed on $S_{12}'$ and $Q$ define the region $efb$ (Fig. \ref{fig:regions-ST}). Because there exist pairs $(\textbf{m},\textbf{n})$ for which Poisson's ratio is positive and pairs for which $\nu$ is negative, elastic materials belonging to these two regions are auxetics.\newline     
iv) If $S_{12}'\left(s_{2},s_{3}\right)<0$, $Q\left(s_{2},s_{3}\right)>0$, then $\nu/e\leq -Q/4$. The inequalities imposed on $S_{12}'$ and $Q$ define the region $ache$ (Fig. \ref{fig:regions-ST}). Because there exist pairs $(\textbf{m},\textbf{n})$ for which Poisson's ratio is positive and pairs for which $\nu$ is negative, elastic materials belonging to these two regions are auxetics.\newline     

II. Consider the pair of unit vectors $\textbf{m}=\left\langle011\right\rangle$, $\textbf{n}=\left\langle100\right\rangle$. In this case $P(\textbf{m},\textbf{n})$ reaches its minimal value 0, and $\nu=-s_{12}$, $e=s_{11}$.\newline
If $S_{12}'\left(s_{2},s_{3}\right)>0$, $Q\left(s_{2},s_{3}\right)>0$, then $\nu<0$. For $s_{3}<0$ the inequalities imposed on $S_{12}'$ and $Q$ define  the region $aed$, whereas for $s_{3}>0$ -- the region $ebd$ of ST (Fig. \ref{fig:regions-ST}). Because for all pairs $(\textbf{m},\textbf{n})$ Poisson's ratio is negative, elastic materials belonging to these two regions are \emph{completely} auxetics.\newline
ii) If $S_{12}'\left(s_{2},s_{3}\right)<0$, $Q\left(s_{2},s_{3}\right)<0$, then $\nu\geq-s_{12}$. For $s_{3}>0$ the inequalities imposed on $S_{12}'$ and $Q$ define the region $cfe$, and for $s_{3}<0$ -- $ehc$ (Fig. \ref{fig:regions-ST}). Because for all pairs $(\textbf{m},\textbf{n})$ Poisson's ratio is positive, elastic materials belonging to these two regions are \emph{non} auxetics.\newline   
iii) If $S_{12}'\left(s_{2},s_{3}\right)>0$, $Q\left(s_{2},s_{3}\right)<0$, then $\nu/e\geq -s_{12}$. For $s_{3}>0$ the inequalities imposed on $S_{12}'$ and $Q$ define the region $efb$ (Fig. \ref{fig:regions-ST}). Because there exist pairs $(\textbf{m},\textbf{n})$ for which Poisson's ratio is positive and other pairs for which $\nu$ is negative, elastic materials belonging to these two regions are auxetics.\newline     
iv) If $S_{12}'\left(s_{2},s_{3}\right)<0$, $Q\left(s_{2},s_{3}\right)>0$, then $\nu/e\geq s_{12}$. The inequalities imposed on $S_{12}'$ and $Q$ define the region $ache$ (Fig. \ref{fig:regions-ST}). Because there exist pairs $(\textbf{m},\textbf{n})$ for which Poisson's ratio is positive and other pairs for which $\nu$ is negative, elastic materials belonging to these two regions are auxetics.    

We conclude that criteria of Ting and Barnett in the unique way divide ST into regions having definite auxetic properties. These regions are displayed in Fig. \ref{fig:regions-ST}. Note that $\nu$ can be isotropic only for complete auxetics. For auxetics Poisson's ratio have to be anisotropic.

\begin{figure}[htpb]
	\centering
		\includegraphics[bb=0 0 300 240, clip]{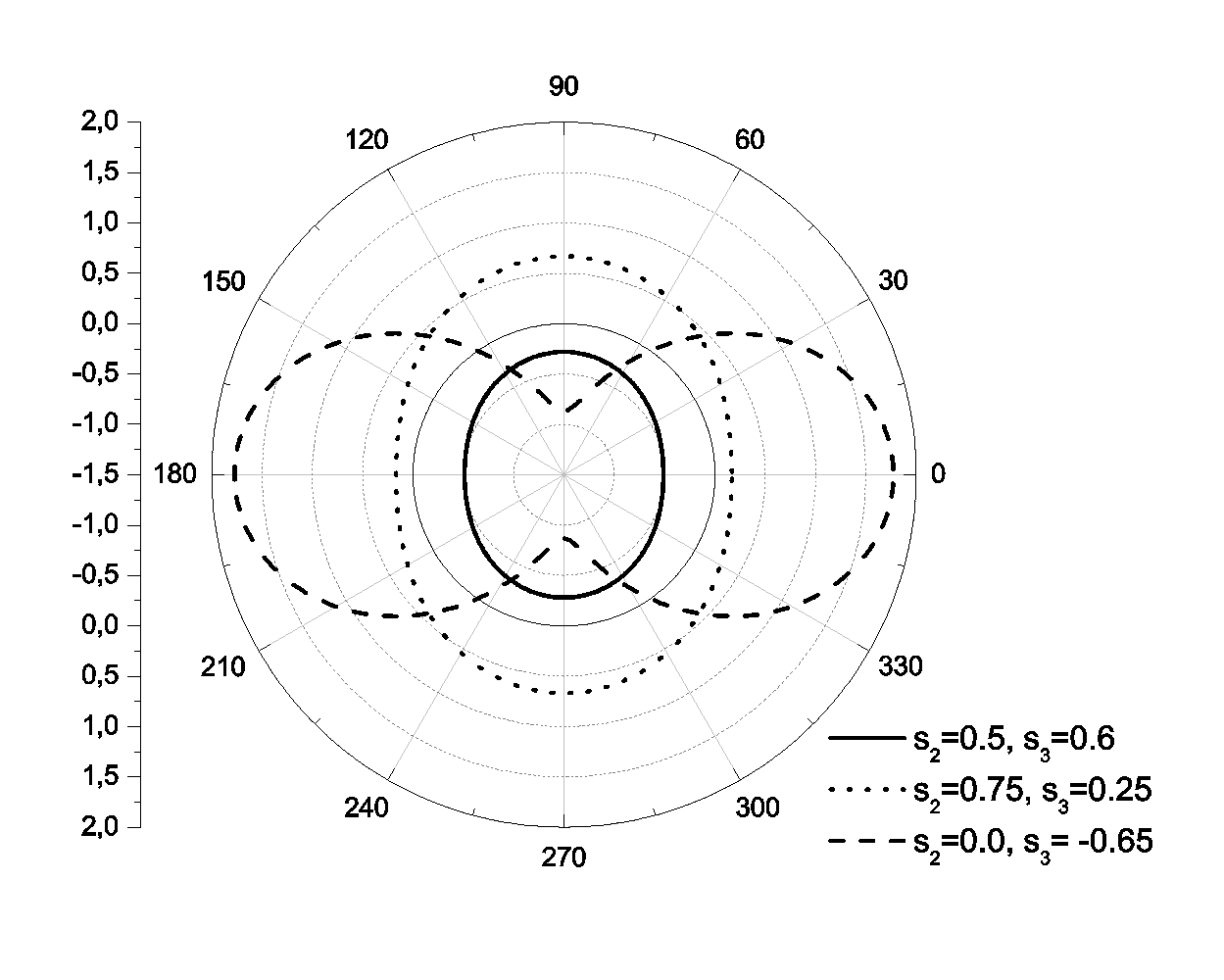}
	\caption{Dependence of $\nu(\textbf{m},\textbf{n};s_{2},s_{3})$ on directions of \textbf{m} perpendicular to $\textbf{n}=[110]$ for several points belonging to characteristic regions of ST }
	\label{fig:nu-anisotropy}
\end{figure}

Finally, we present results of our calculation of Poisson's ratio, which confirm results described in previous sections of our paper. For $\textbf{n}=<110>$, the dependence of $\nu$ on directions $\textbf{m}$ for points belonging to the characteristic regions of ST is presented in Fig. \ref{fig:nu-anisotropy}. The solid line represents complete auxetics (the region \emph{ebd} in Fig. \ref{fig:regions-ST}). The dotted line represents complete nonauxetics (the region \emph{cfe}), and the broken line -- (non-complete) auxetics (the region \emph{ache}). For them Poisson's ratio has to be anisotropic  

\begin{figure}[htpb]
	\centering
		\includegraphics[bb=0 0 298 232, clip]{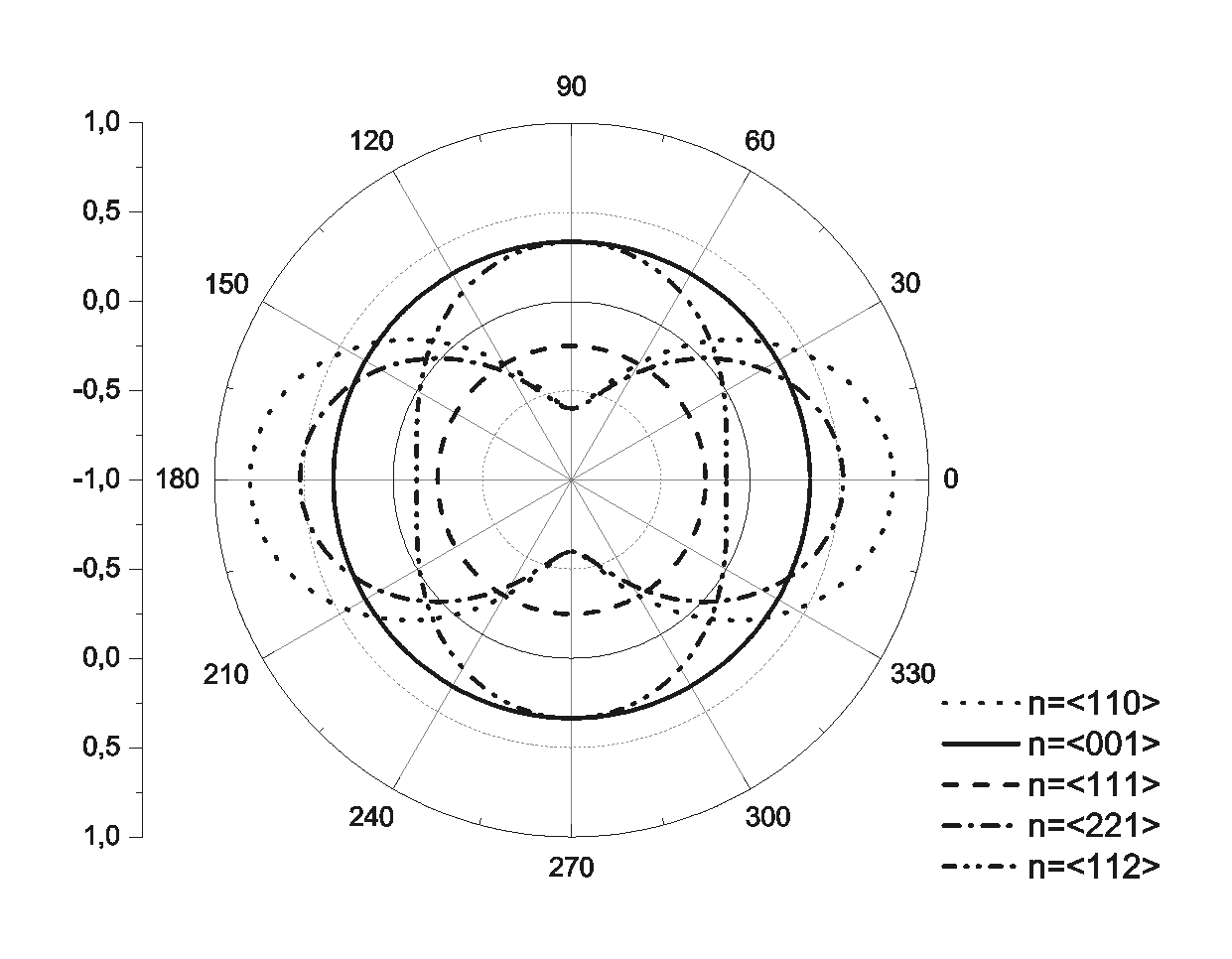}
	\caption{Dependence of $\nu$ on directions of \textbf{m} perpendicular to several directions $\textbf{n}$ for ST point $s_{2}=-0.2$, $s_{3}=-0.65$ belonging to $ache$ -- the auxetic region of ST}
	\label{fig:nu-on-ST}
\end{figure}

Anisotropy properties of Poisson's ratio for several vectors $\textbf{n}$ for the point $s\equiv \left(s_{2}=-0.2,\,s_{3}=-0.65\right)\in ache$ region of ST, are shown in Fig. \ref{fig:nu-on-ST}.
\section*{Acknowledgements}We express our gratitude to A.G. Every for making the authors aware of the paper by Ting and Barnett, and to T.C.T. Ting for sending us a reprint of it. We would like to Professors R. Lakes and A. Bra\'{n}ka for critical reading of the manuscript.  
\appendix
\section{}
In the Appendix we write down the explicit formulas for Young's, shear modules and for Poisson's ratio of cubic media in terms of components of the stiffness and compliance tensors.
\begin{equation}
\frac{1}{E(\textbf{n})}=\frac{1}{3(C_{11}+2C_{12})}-\frac{1-3p(\textbf{n})}{3(C_{11}-C_{12})}+\frac{1-p(\textbf{n})}{2C_{44}},
\label{eq:a1}
\end{equation}

\begin{equation}
\frac{1}{4G(\textbf{m},\textbf{n})}=\frac{P(\textbf{m},\textbf{n})}{\left(C_{11}-C_{12}\right)}+\frac{1-2P(\textbf{m},\textbf{n})}{4C_{44}},
\label{eq:a2}
\end{equation}

\begin{equation}
\nu(\textbf{m},\textbf{n})=E(\textbf{n})\frac{(C_{11}+2C_{12})(C_{11}-C_{12}-2C_{44})P(\textbf{m},\textbf{n})+2C_{44}C_{12}}{2(C_{11}+2C_{12})(C_{11}-C_{12})C_{44}}
\label{eq:a3}
\end{equation}
\begin{equation}
\frac{1}{E(\textbf{n})}=S_{12}+\frac{1}{2}S_{44}+\left(S_{11}-S_{12}-\frac{1}{2}S_{44}\right)p(\textbf{n}),
\label{eq:4}
\end{equation}

\begin{equation}
\frac{1}{4G(\textbf{m},\textbf{n})}=\frac{1}{4}S_{44}+\left(S_{11}-S_{12}-\frac{1}{2}S_{44}\right)P(\textbf{m},\textbf{n}),
\label{eq:a5}
\end{equation}

\begin{equation}
\nu(\textbf{m},\textbf{n}) =-E(\textbf{n})\left[S_{12}+(S_{11}-S_{12}-\frac{1}{2}S_{44})P(\textbf{m},\textbf{n})\right].
\label{eq:a6}
\end{equation}

When $S_{44}=2\left(C_{11}-C_{12}\right)$ or $S_{44}=\left(C_{11}-C_{12}\right)/2$ one deals with an isotropic medium. If these relations are fulfilled 
from Eqs. (\ref{eq:a1}-\ref{eq:a6}) one obtains expressions for mechanical characteristics of isotropic media 
\begin{eqnarray}
	E^{-1}=\frac{C_{11}+C_{12}}{\left(C_{11}+2C_{12}\right)\left(C_{11}-C_{12}\right)}=S_{11},\nonumber \\
	G=\frac{1}{2}\left(C_{11}-C_{12}\right)=\frac{1}{2\left(S_{11}-S_{12}\right)},\nonumber \\
	\nu=-ES_{12}=\frac{C_{12}}{C_{11}+C_{12}}=-\frac{S_{12}}{S_{11}}.\nonumber
\end{eqnarray}

The knowledge of the components of the compliance tensor of cubic media allows one to find the corresponding values of parameters $s_{l}\; (l=1,2,3)$ 
\[
s_1=-\frac{S_{44}S_{12}+S_{44}S_{11}-4S_{12}^2+2S_{12}S_{11}+2S_{11}^2}
{(2S_{12}+S_{11})(S_{12}-S_{11})S_{44}},
\]
\[
s_2=\frac{S_{44}S_{12}+S_{44}S_{11}+2S_{12}^2-S_{12}S_{11}-S_{11}^2}
{S_{44}S_{12}+S_{44}S_{11}-4S_{12}^2+2S_{12}S_{11}+2S_{11}^2},
\]
\[
s_3=\frac{(2S_{12}+S_{11})(2S_{12}+S_{44}-2S_{11})}
{S_{44}S_{12}+S_{44}S_{11}-4S_{12}^2+2S_{12}S_{11}+2S_{11}^2}.
\]

\end{document}